\documentclass[preprint,floats,eqsecnum,aps,showpacs,showkeys]{revtex4}
\usepackage{epsfig}
\usepackage{color}
\begin{document}

\newcommand{\be}{\begin{equation}}
\newcommand{\ee}{\end{equation}}
\newcommand{\bea}{\begin{eqnarray}}
\newcommand{\eea}{\end{eqnarray}}

\newcommand{\SG}{\sigma}
\newcommand{\ep}{i\varepsilon}
\newcommand{\nn}{\nonumber}
\newcommand{\al}{\alpha}
\newcommand{\psie}{\hat{\psi}}
\newcommand{\Ae}{\hat{A}}
\newcommand{\om}{\omega}
\newcommand{\ArabicAgain}{\renewcommand{\thechapter}{\arabic{chapter}}}
\newcommand{\dl}{\delta}

%%%%%%%%%%%%%%%%%%%%%%%%%%%%%%%%%%%%%%%%%%%%%%%%%%%%%%%%%%%%%%%%%%%%%%%

\title{Lattice data inspired but  Minkowski space calculated  QCD fundamental propagator}

\author{Vladimir ~\v{S}auli}
\affiliation{Department of Theoretical Physics, INP Rez, CAS, Czech Republic}

\begin{abstract}
We study the Dyson-Schwinger equation for the quark propagator in Minkowski space. In order to have analytical behaviour  at the timelike axis of momenta under control, we use the  Stieltjes and the Hilbert transformation for the interaction kernels and discuss the solution from the perspective of these  transformations. In addition, a lattice fit for the gluon propagator and approximation for quark-gluon vertex are employed, and within the model the quark propagator is obtained through the solution of Dyson-Schwinger equation in Minkowski space.  The resulting propagators in all studied cases do not  show up  particle like pole and production thresholds. Instead of,  the quark propagator satisfies Hilbert transformation and the associated dynamical mass function becomes complex without a presence of particle like branch point. 
\end{abstract}    


\maketitle

%-------------------------------------------------------------------------------------------------------
%-----------------------------------------------------------------------------------------------------------------------

\section{Introduction}

The QCD Green's functions (GFs)  can be used as a building blocks when for calculating the properties of hadrons 
\cite{BCCELRT2012,MATA1999,ROBWIL1994,BHATAN2004,CLOROB2013} at  arbitrary scale.
In the last decade considerable progress in the lattice evaluation of QCD fundamental GFS  have been made \cite{CUME2007,BIMS2007,OLSI2009,KLSW2005}, being in a reasonable numerical agreement with the calculation performed in the continuous framework of Dyson-Schwinger equations (DSEs) \cite{ABIPA2008,BIPA2009,FIS2006,AGPA2011,ABICAPA2013,HOWIAL2013}.  The DSEs are the integral equations, i.e. they are based on the utilizing of continuous space time, tacitly, up to a few exceptions,  the DSEs studies were performed in the Euclidean space, establishing thus a meaningful comparison with the lattice data. Also the most of the DSEs solutions we restricted to Landau gauge, which is the only one among more general covariant gauges,  achieved numerically in the lattice.  Mainly due to this reason, the lattice inspired is formally restricted to this gauge only, while the main focus on the analytical structure and related points are quite general and independent on any gauge choice.

In  recent agreement with the lattice data,  the gluon propagator provides infrared finite, however non vanishing solution. 
This phenomena has been  explained in term of non-Abelian Schwinger mechanism in Background Field method 
\cite{BIPA2009}. Quite related, but only recently clarified point is  the understanding of significant role of the dressed quark-gluon vertex for a  correct description of dynamical chiral symmetry breaking in QCD.  
There is a wide consensus that without infrared enhancement of quark-gluon vertex the chiral symmetry breaking does not occur.
 
On the other side many questions remain unsolved, e.g.  the relation between confinement of fundamental QCD object -the quarks and gluons- and the dynamical symmetry breaking is not yet completely understand. Recall, the confinement is conventionally understand  from the study of Wilsonian loops,
 and various confining  potentials are calculated  in the static limit of infinitely heavy quarks.  In this paper, we  argue that  chiral symmetry breaking an the confinement of light quarks, albeit slightly differ from the conventional wisdom, actually happen simultaneously, which fact is directly observed  when the DSEs are considered and solved  in  our real world Minkowski space time. 
We argue that due to the strong coupling at low energy, the QCD quantum loops lead to the complex quark and gluon dynamical mass functions, without usual appearance of a real valued branch points.

The equations for QCD propagators represent a part of  SDEs system. For practical purpose of solution it is convenient to 
express them  in the the form of  irreducible GFs, i.e. in terms of the selfenergies, polarizations and proper vertices. Then SDE equation for the quark propagator reads
\be  \label{sigma}
S^{-1}(p)=\not p-m_0-\Sigma(p) \, ,
\ee
where selfenergy function $\Sigma$  therefore completely determines the analytical property of the propagator. 

Contrary to the solution of SDEs performed in the Euclidean space, the numerical solution in Minkowski spacetime is not a well defined problem,
unless one does not specify analytical boundary condition properly. The problem is that we are looking for solution in terms of (generally non-unique) tempered distribution. It turns that a "physical" solution is given as a limit of complex multivaluable functions at the region we expect the function has an analytical cut. Obviously, as the computer works with  columns of the numerical data and a desired analytical property is not directly implemented, the numerical search easily fails  for almost any admissible method.

Let us recall here, that the Euclidean space quark and gluon propagators have  been quite accurately approximated by the meromorphic functions with complex conjugated poles \cite{STI1996,CUME2008,ABIPA2008,FIMAPA2009,DOS2013,GRA2010,MAMEOL2011,BBBSLMPPR2011,OLBI2011,DUGU2011,GRA2012,WIHUAL2013}. In the first part of presented paper we study analytical structure of a simple correlator made out of the propagators with various analytical properties. We concern on the appearance of the lowest real branch point assuming the propagator satisfy Hilbert or Stieltjes transformation which automatically includes the case of the complex conjugated pole as well. In most of the cases  a naive Wick rotation is invalid and the integration in Minkowski momentum spacetime must be carefully reconsidered. In the second part of the paper, the DSE for the quark propagator is solved numerically in Minkowski momentum space assuming the kernel has a special analytical properties discussed in the preceding Section.  The Section IV provides the numerical solution for a lattice inspired kernel of the Schwinger-Dyson equation. 

% --------------------------------------------------------------------------------------------------------------------------

\section{ Perequisities from scalar correlator in Minkowski space}

In nonconfinig theory and in the case of stable particle, the selfenergy $\Sigma$ in Eq. (\ref{sigma}) should be real bellow the threshold, allowing thus appearance of a real pole in $S$, which uniquely corresponds with
a free propagation of simple field excitation mode -the particle.
% funkce muze mit vic polu, ktere semohou vzajemne vyzrat v s-matici, nicmene zbyvajici poly opet koresponduji        
More then 50 years ago, hereby for a non-confining theory, the analytical structure of perturbative Feynman graphs have been investigated 
\cite{KSW1958,LANDAU1959,CUT1960}. Especially, it is well known, that two point correlators satisfy the integral representation, which allows  write down
a simple relation between its imaginary and  real part. The following spectral representation 
\be \label{spectrality}
\int d\alpha \frac{\rho(\alpha)}{p^2-\alpha+i\epsilon}
\ee
was found to be a very useful tool for solving DSEs nonperturbatively in many models without confinement phenomena, however it usually fails, when trying to extend its use  for a description of the dynamical  chiral symmetry breaking phenomena and QCD bound states, e.g. the pion. Quite interestingly, it was found that the residue of propagator where vanishing and the Lehman spectral weight function turned to be sign-changing when approaching critical value of couplings. In other words, for increasing coupling, the associated  particle modes have tendencies to disappear from the spectrum and the  probability interpretation of spectral density is lost.  For a historical  suggestions for using spectral representation in nonperturbative context see \cite{SAL1963,STRA1964,DELWES1977A,DELWES1977B,DEL1979,JMC1982}, for an actual numerical solutions see \cite{SAU1,SAU2}.

 In  QCD the only freely moving objects are the bound states - the hadrons. Thus the correlators made from the hadronic currents may obey various sum rules well based on the  dispersion relations \cite{SVZ1979,SHI1998,KOLKHO2001,BRAUN1998,GUOK2010}. 
In contrary to hadrons, the colored objects like quarks are not observed, thus the 
proof or reliable evidence for the absence of singularities associated with quark (gluon)  productions would certainly comply with 
confinement of quarks (and gluons) quanta.  This has been translated into the lattice language of quark Wilsonian loop, where the are law represents indirect search for a (an absence of) quark-antiquark threshold \cite{WILSON}.  A more direct observation based on the study of QCD GFs study in timelike region of Minkowski space-time is barely lacking.

As a first step we write down the Hilbert-Stjieltjes transformation for the  propagators, as the second step we rewrite this transformation into the form of unbounded Feynman and Dyson integral representations, which will allow us to calculate  loop integrals in Minkowski space directly.

In order to understand the feedback of various analytical properties  on the  selfenergy  $\Sigma$ (\ref{sigma})
 we will consider the  following scalar correlator 
\be     \label{corelator}
\Pi(p^2)=i\int \frac{d^4l}{(2\pi)^4} G_1(l)G_2(l-k) \, ,
\ee
where the functions $G_{1,2}$ mimic propagators of quarks or gluons in deep infrared. For this purpose we will consider the function $G$, which
obey Stieltjes transformation in the forms:
\be 
G_{{\cal S}_+}(p)=\int_0^{\infty} d x \frac{\sigma(x))}{p^2-x+i\epsilon} \, ; \,G_{{\cal S}_-}(p)=G_{{\cal S}_+}^+(p) \, ; \,
G_{\cal S}(p)= \int_0^{\infty} dx \frac{\sigma(x)}{p^2-x}
\ee
noting the spectral representation belongs to  ${\cal S}_+ $ function. In addition we will also consider the function 
which are images of Hilbert transformation, e.g. : 
\be  \label{hilbert}
G_{{\cal H}_+}(p)=\int_{-\infty}^{\infty} dx\frac{\sigma(x)}{p^2-x+i\epsilon} \, ; \,G_{{\cal H}_-}(p)=G_{{\cal H}_+}^+(p) \, ; \,
G_{{\cal H}}(p)=\int_{-\infty}^{\infty} dx \frac{\sigma(x)}{p^2-x} \, .
\ee

Considered spaces  classify the functions with respect of the position of the singularities in the whole complex plane, and also with respect to position of possible branch point located at the real axis. Recall for instance that purely meromorphic functions with complex conjugated poles suggested in 
 \cite{MATA1999,BHATAN2004,GRA2010,OLBI2011,DUGU2011,GRA2012,DOS2013,CLOROB2013} belongs entirely to ${\cal H}$.
, i.e to the function which can be expresses through the unbounded real valued principal value integral Hilbert transformation (to name the transformation we follow the book \cite{ERD}).    
It is a matter of the fact that the function with  singularities located simultaneously at upper and lower half plane of complex
do not satisfy representation of the form (\ref{spectrality}), thus it must either from $H$ or  $S$, the later implies the presence of positive branch point at the real axis.  Actually one can easily check that  the presence of positive (negative) epsilon ensures that all singularities are located at the lower (upper) plane of complex $p^2$ plane.
Note trivially, the meromorphic function can be written as a combination of $H_+$ and $H_-$ or $S_+$ and $S_-$ because of the identity
\be  \label{iden4}
G_H(p)=P. \int_{-\infty}^{\infty}\frac{d x \, \sigma(x)}{p^2-x}=\int_{-\infty}^{\infty} \frac{ dx\, \sigma(x)}{p^2-x+\ep}+\int_{-\infty}^{\infty}
 \frac{dx\, \sigma(x)}{p^2-x-\ep} \, .    
\ee
As we will discuss in the next section the numerical solutions of quark DSEs actually suggests that the quark propagator is the linear combination
of the functions which belongs to  $H_+$ and $H_-$. It is plane to say, that one naturally assume that they contribute asymmetrically as we know that only 
$H^+$ ($S^+$) dominates the spacelike ultraviolet as a consequence of asymptotic freedom.

To illustrate of above saying, let us write down a simple example:
\be \label{exam}
\int_a^{\infty} dx \frac{(a-x)g(x,a,b)}{p^2-x}=g(p^2,a,b)\left[\frac {\pi b }{2}+\frac{(p^2-a)}{2} ln\left(\frac{p^2-a}{b}\right)^2\right] 
\ee
being a case of $S$ transformation.
Whilst the Hilbert transformation of the same weight function is meromorphic 
\be
\int_{-\infty}^{\infty} dx \frac{(a-x)g(x,a,b)}{p^2-x}=\pi b g(p^2,a,b) \, ,
\ee
and where we have used the shorthand notation  for the "Gribov" propagator $g(x;a,b)=[(x-a)^2+b^2]^{-1}$ with cpx. conjugated poles
at $a\pm i b$.

From  discussion above it follows that most general case which we will need to consider is the  sum of 
Feynman-Hilbert and Dyson-Hilbert representations for considered propagator 
\be \label{STE2}
G(p^2)=\int dx \frac{\sigma_+(x)}{p^2-x+\ep}+\int dx \frac{\sigma_-(x)}{p^2-x-\ep} \, .
\ee
By construction any function given by solely $P$ integral is included by the virtue of the identity (\ref{iden4}).

Before doing so we mention a well known and easily derivable from the well known:

For the convolution of two functions from two function from $S$ we  get $\Pi=0$ everywhere.

For the product of two function from $S_+$ we get the function from $S_+$ again, which is nothing else but the regular case of perturbation 
theory Feynman diagram. It can be written  in the form of well known dispersion relation
\bea\label{dis}
\Pi(p^2)&=&\int_0^{\infty} dx \frac{\rho_{\pi}(x)}{p^2-x+\ep}
\nn \\
\rho_{\pi}(x)&=&\int_0^{\infty} \frac{d y d z}{8\pi^2} \sigma_{+1}(y)\sigma_{+2}(z) \frac{\sqrt{(x-y-z)^2-4yz}}{x} \Theta(x-(y^{1/2}+z^{1/2})^2)
\eea
where the function  $\sigma_{+1}(y)$ and $\sigma_{+2}(y)$ are the weights of the images $G_1$ and $G_2$. The derivation is the
textbook example for one  loop Feynman diagram (see \cite{SAULI} for its regularized form).

\begin{figure}\label{xy}
\centerline{\epsfig{figure=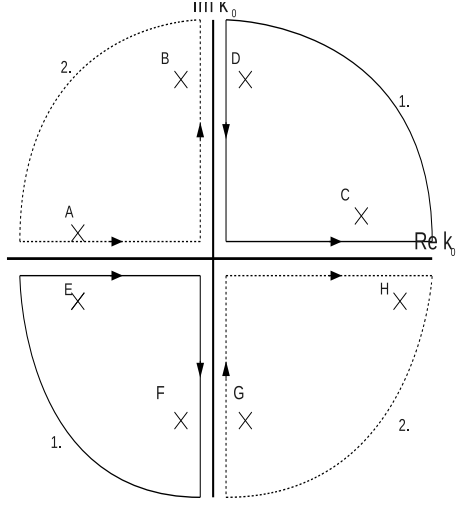,width=7truecm,height=7truecm,angle=0}}
\caption[caption]{Integration contours in complex momentum plane used for the integration over $l_0$ component.   Contour filling the first and the third quadrant is used for the  first line while the mirror symmetric contour was used to integrate the fourth line of Eq. (\ref{rozpis}). Position of singularities are shown for fixed $\omega$, the ones close to the real axis (A,C,E,H) correspond to positive value of $\vec{l}^2+\omega$, while the ones closed to Im axis (B,D,F,G) appear when $\vec{l}^2+\omega$  is negative. i.e. for Hilbert transformations only. }
\end{figure}

For the convolution of the function from  $S_-$ one gets
\be 
\Pi(p^2)=-\int dx \frac{\rho_{\pi}(x)}{p^2-x-\ep} \, ,
\ee
with the same $\rho_{\pi}$ as previously and  where the sign is a consequence of the integration over $k_0$ in momentum space (one can use the mirror symmetric contour displayed at the  fig. \ref{xy}, or one can simply conjugate previously considered $\Pi$ multiplied by $1/i$ before.
 
Let us study the most general case (\ref{STE2}).
In order to get sense to the product of the functionals we rewrite 
the correlator (\ref{corelator}) into the form where all products belong to same "type" of distributions 
\bea\label{rozpis}
\Pi(p^2)&=&i \int\frac{d^4l}{(2\pi)^4}G_1 G_2
\nn \\
G_1G_2&=&\left[\int \frac{d\om_1 \sigma_{+1}(\om_1)}{l^2-\om_1+ \ep}+\int \frac{d\om_1 \sigma_{-1}(\om_1)}{l^2-\om_1- \ep}\right]
\nn \\
&&\left[\int \frac{d\om_2 \sigma_{+2}(\om_2)}{(l-p)^2-\om_2+ \ep}+\int \frac{d\om_2 \sigma_{-2}(\om_2)}{(l-p)^2-\om_2- \ep}\right]
\nn \\
\Pi(p^2)&=&i \int d\om_1 \int d\om_2 \int \frac{d^4l}{(2\pi)^4}
\nn \\
&&\left[ \frac{\sigma_{+1}(\om_1)\sigma_{+2}(\om_2) }{(l^2-\om_1+ \ep)((l-p)^2-\om_2+ \ep)}+
 \frac{ \sigma_{-1}(\om_1)\sigma_{-2}(\om_2)}{(l^2-\om_1- \ep)((l-p)^2-\om_2- \ep)} \right.
\nn \\
&+&\left. \frac{\sigma_{+1}(\om_1)\sigma_{-2}(\om_2) }{(l^2-\om_1+ \ep)((l-p)^2-\om_2- \ep)}+
 \frac{ \sigma_{-1}(\om_1)\sigma_{+2}(\om_2)}{(l^2-\om_1- \ep)((l-p)^2-\om_2+ \ep)} \right]
\eea

If $G_i$ are from $S$, then the  first two terms correspond with already discussed cases. These two terms then have  the lowest branch point located at real positive  semi-axis.

The rest can be easily evaluated by using the following identity 
\be
\frac{1}{x-a\pm \ep}= P. \frac{1}{x-a}\mp i\pi\delta(x-a) \, .
\ee
Using a shorthand notations  $\delta_{1}=\delta(l^2-\omega_1)$, $\delta_2=\delta((l-p)^2-\omega_2)$ and analogously pro principal parts, then   
the last line can be written as:
\bea
&&\sigma_{+1}(\om_1)\sigma_{-2}(\om_2)\left[ P_1 P_2-i\pi\delta_1 P_2+i\pi\delta_2 P_1+\pi^2\delta_1\delta_2\right]
\nn \\
&&\sigma_{-1}(\om_1)\sigma_{+2}(\om_2)\left[ P_1 P_2+i\pi\delta_1 P_2-i\pi\delta_2 P_1+\pi^2\delta_1\delta_2\right]
\eea

Now for the function from $S$'s, the first  terms at each lines are  zero separately, the second and the third terms produces purely real pieces in the final result for the correlation function and  last terms at each lines are exactly what the Cutkosky rule would give for absorptive part, however with opposite sign and  for the continuous mass there. Explicitly
\be
i\int_0^{\infty} \frac{d \om_1 d \om_2 }{8\pi} \sigma_+^1(\om_1)\sigma_-^2(\om_2) \frac{\sqrt{(p^2-\om_1-\om_2)^2-4\om_1\om_2}}{p^2} 
\Theta(p^2-(\om_1^{1/2}+\om_2^{1/2})^2)
\ee
for the first line and similar expression is valid for the last  term of the second line (with an appropriate weights). Again one can conclude that  for the inner momentum integral one gets the usual perturbative branch point $\om_1^{1/2}+\om_2^{1/2}$, which is further smeared by  the integrations  over the weights $\sigma$'s. 
While a detailed property of the branch point does depend on the   Stieltjes weights, the lowest branch point remains bounded at timelike regime ( it is equal to $4a^2$ for the example (\ref{exam})). In words, the correlation functions remain real for 
a spacelike argument for the case of Stieltjes transformation, no matter whether the propagators $G$ involves complex conjugated poles or not.   
 Note for completeness, that for the all weight functions 
identical, the most contributions exactly cancel against each other, and  trivial identity remains: $P_1 P_2=P_1 P_2$.

As a last but not at least example let us consider the function which involves also Hilbert images. It will dramatically changed the situations when compared to the functions represented by the functions solely from $S$. We will consider only  special case when the both functions are from  $H$ space,
 then after some effort, one can arrive into a relatively  simple expression 
\bea
\Pi_H(p^2)&=&\int_{-\infty}^{\infty}d\om_1  \int_{-\infty}^{\infty}d\om_2 \sigma_1(\om_1) \sigma_2(\om_2) K(p^2,\om_1,\om_2)
\nn \\
K_H(p^2;\om_1,\om_2)&=&\frac{-i}{16\pi }\left[(1+\frac{\omega_1-\omega_2}{p^2})\Theta(-\omega_1)+(1-\frac{\omega_1-\omega_2}{p^2})\Theta(-\omega_2)\right] \, ,
\eea
which is valid even for spacelike $p^2$. The result is obviously nonzero for general $\sigma_{1,2}$ and the resulting nontrivial pieces arise from the  region where principal integral turns to be an usual regular integral. The result is obviously complex everywhere and also the combination of the function from $H_+,H_-$ and $ H$ spaces with any other tempered distribution  considered in this section produces the correlator which is complex in the spacelike region of momenta as well.

Stress several things here, first of all it is impossible   to have  QCD propagators entirely from $H$'s, noting the variable $\omega $ is linearly appearing in the numerator and the appropriate divergences would lead to a deep contradictions with perturbative QCD and the best QCD experiments. It is pertubatively true, that  when one starts with all $G$'s form $S^+$ then one finishes with the function from $S^+$ again. This is a basic ingredient of Perturbation Theory Integral Representation derived for more complicated Feynman integrals by Nakanashi \cite{Nakanishi}.    
 However,the  function from $H$  can be generated nonperturbatively, giving important contribution at the  infrared  $p \simeq \Lambda$ and being thus important for confinement. This is in fact the case of Minkowski space-time  study \cite{KAYMUN1996}, wherein the full quark propagator instead of spectral representation satisfies a Hilbert transformation.

In the all above examples  we have silently assumed that the weight functions are  real. At the end we we shall mention some attempt to use complex spectral function in the context of the so called spectral quark model \cite{ArrBro2003}.
In the paper \cite{ArrBro2003} a many low energy strong QCD hadronic quantities and form form factors have been formally evaluated in the terms of the integral representation with purely imaginary spectral functions as a consequence of phenomenologically guessed
contour of integration. One should worry with  uniqueness of such prescription since the $\rho$ usually represents the discontinuity of the functions at the cut and thus is not well defined if $\rho$ has a cut there as well.  Let us recall that there exist formally infinity number of  imaginary spectral  $\rho$'s which makes a real $G$ because of the identity:
\be 
G(p^2)=\frac{i}{2\pi}\left(\int_{M}^{\infty} dx \frac{G(x)}{p^2-x+\ep}-\int_{M}^{\infty} dx \frac{G(x)}{p^2-x-\ep}\right) \, .    
\ee
which is an exact property of the quark propagator maintained in the spectral quark models (however here $G$ is bounded bellow by $M$). Irrespective of possible phenomenological success, we expect a concept 
of complex $\rho$ is mismatching  a  desired usefulness and usual understanding of dispersion relation.

\section{Rainbow-Ladder quark DSE in Minkowski space within Hilbert, Stieltjes and Khallen-Lehmann representation for the kernel}

In the paper \cite{KAYMUN1996} the confining model based on the model gluon  propagator (in position space)
\be \label{choice}
G^{\mu\nu}(x)=i\mu^2 g^{\mu\nu}
\ee 
has been considered. This  study has been  performed directly in the Minkowski space avoiding thus necessity of analytical 
continuation from Euclidean space. Very interestingly, the authors finished with the solution which is given by Hilbert transformation
$H$ (\ref{hilbert}) and not a Feynman propagator (Stieltjes $S_+$ transformation).  McKay-Munczek quark propagator turns to be a  real function for 
all real momenta $p^2$ with the following infrared singularity
\be
PP. \frac{e^{-m^2/2\mu^2}}{p^2} \, .
\ee    
That a purely imaginary kernel can give a purely real solution is not a big surprise, as long as one is working with distribution (\ref{choice}).
Obviously the effective kernel is chosen to be purely imaginary, for complex  parameter $\mu^2$  we can get a complex solutions as well.
 
In this section we study the quark DSE numerically in momentum Minkowski space. We begin by developing the model in the ladder-rainbow approximation with the kernels, which will be approximated by infrared enhanced function which has or has not complex conjugated singularities.
Instead by taking some ad hoc Ansatz we have consider a suited combination of the functions from spaces $S$ and $H$. We have use the same weight functions for  Stieltjes and Hermite transformation differing thus in analytical properties of constructed the kernel. However we leave the quark propagator undetermined and expect the solution should fall into some of the classes $S,H$, either the combination is admitted as well. We restrict  to the presentation of three simple combinations, noting here the other possible combinations do not differ qualitatively.    

For completeness we review basic ingredients for the quark propagator DSE, which  reads
\bea  \label{system}
S^{-1}(p)&=&\not p -\Sigma(p)=A(p)\not p- B(p)
\\ \nn
\Sigma(p)&=&ig\int\frac{d^4k}{(2\pi^4)} \Gamma^{\mu}(q,p,k) S(k)\gamma^{\nu} G_{\mu\nu}(q)
\\ \nn
G_{\mu\nu}(k)&=&=\left(g_{\mu\nu}-\frac{k_{\mu}k_{\nu}}{k^2}\right)D(k^2) \, ,
\eea
where the usual conventions were followed: $S$ stands for the quark propagator, which up to a tiny electroweak corrections is well described by 
two scalar function $A,B$. $G$ is a a gluon propagator and  $\Gamma$ stands for quark-gluon vertex, which turns to be an important and necessary 
ingredience for a correct description of confinement and chiral symmetry breaking.   
We have suppressed all color and Dirac indices for brevity, explicitly written  

\be 
\Gamma(q,p,k)=gt_{ij}^a\Gamma^{\mu}_{\alpha\beta}(q,p,k); \, \, \Gamma^{\mu\, [{\mbox Ladder}]}_{\alpha\beta}=\gamma^{\mu}_{\alpha\beta}Z
\ee
where $q,p$  are 4-momenta of incoming gluon and quark and $k$ stand for outgoing quark 4-momentum  respectively, $Z$ is a Lorentz scalar in the ladder-rainbow approximation, which we will use  in this Section, i.e.  only a $\gamma$ part out of all twelve component 
 of the vertex $\Gamma$ is considered. Explicitly we get for the function $B$   
\bea \label{defker}
B(p)&=&\frac{Tr}{4}ig\int\frac{d^4k}{(2\pi)^4} \gamma^{\mu} t^a S(k)\gamma^{\nu} t^a G_{\mu\nu}^{[free]}(q)\alpha(q)
\nn \\
&=& i\int \frac{d^4k}{(2\pi)^4} \frac{B(k)}{k^2-B(k^2)} K(q)
\eea
and similarly the equation for $A$ can be derived by projection by $Tr \not 4p$. We will neglect the momentum dependence of $A$, simply noting 
that variation of $A$ is expected to be quite small in the approximation used \cite{ArrBro2003}.
In the last DSE $\alpha$ is an effective running coupling made of the $\gamma$ vertex and gluon  form factor functions..

We will consider three models  defined by the kernel $K$, which is chosen to be either from $S$'s or $H$ spaces.
As a first very conventional choice we will consider the kernel which satisfies usual dispersion relation,
 i.e. the function which is from  $S^+$. It is chosen in a way the kernel $K$ in Eq. (\ref{defker}) is equivalent to the product of two following functions 
\be  \label{KSP}
K=K_{S_+}(l^2)=(8\pi^3)C_{S+} F_{S+}(l^2)G_{S+}(l^2)
\ee
and where $G$ is the function  (\ref{exam}) with positive Feynman $\ep$ included, i.e.
\be
G_+(x)=\int_a^{\infty} dy \frac{(a-y)g(y,a,b)}{x-y+i\epsilon}
\ee
and where $F_{+}$  is a correct continuation of the following exponential 
\be
F(q)=e^{-\frac{\sqrt{-q^2}}{\Lambda}} ; q^2<0 ,
\ee
which  ensures convergence of DSE and finiteness of the integrals. For all 
momenta  it reads
\be
F_{+}(x)=e^{-\frac{\sqrt{-x}}{\Lambda}}\Theta(-x)+e^{i\frac{\sqrt{x}}{\Lambda}}\Theta(-x)
\ee
and the effective constant is taken such that $(8\pi^3)C_{S+} \simeq 1$.

The second model is defined as
\bea \label{KS}
K_{S}(l^2)&=&(8\pi^3)C_S F_{+}(l^2)G_{S}(l^2)
\nn \\
G_{S}(x)&=&\int_a^{\infty} dy (a-y)g(y,a,b)/(x-y) \, 
\eea
i.e. $G$ being from $S$ implies nontrivial admixture of the function from $S^{-}$ as well.

Finally, for the third model we take
\bea \label{KH}
K_{H}(l^2)&=&(8\pi^3)C_H F_{+}(l^2)G_{H}(l^2)
\nn \\
G_{H}(x)&=&\int_{-\infty}^{\infty} dy (a-y)g(y,a,b)/(x-y)=\pi b g(x,a,b)  \, 
\eea
which means that the kernel in addition to $S_+$  involves  a nontrivial admixture of the functions from $H_+$ and $H_-$.
The  parameters $a,b$ are taken to be equal $a=b=\Lambda^2$ for simplicity. 

Let as stress once more, the second and the third kernels have complex conjugated singularities, while the first one is a conventional Feynman representation, which is singularity free at the upper half  of  the first Riemann sheet. There are branch points associated  with square root in the exponential and there is cut associated with discontinuity along a real axis. Furthermore, a usual logarithmical cut is presented in the first two cases, this cut starts at $x=a$ and goes to positive infinity, associated discontinuity of the full kernel appears explicitly in the first kernel only, 
because of $\ep$ presence. For those who are interested, the  integral representations are easily derivable for the full kernels $K$,
 the example for $K_{S^+}$ is written in the Appendix $A$ for completeness. 

We allow the quark propagator is complex everywhere, assuming that the  imaginary part  will be trivial if the numerical solution of DSE admits.
Some of the details of numerics are written in the Appendix B. As a first, we present the numerical solution for the dynamical mass function 
at spacelike region of momenta, i.e. in the domain where all the Euclidean space results are usually presented. In Fig. 1 the results for small value of quark current mass $ m_0=0.01\Lambda $  are presented for  the three above considered cases.
The letters $H,S$ and $S_+$ label the model where the function $G$ belongs to a given spaces introduced in the previous section and hence to the model defined by the kernel (\ref{KH},\ref{KS}) and (\ref{KSP}) respectively.
The appropriate effective coupling was chosen to be $C=1/90$ for the $S_+$ and $S$ model, while in order to achieve better convergence 
 $C=1/(\pi 90)$ for the third model (the models with pure meromorphic functions are less stable for larger coupling).   
The same is shown in Fig. 2 but for a larger current quark
 mass $m/\Lambda=0.5$ . Thus the first case can mimic light $u,d$ quarks and the second case can correspond with -s quark- mass
 if one identifies $\Lambda$ with QCD scale  $\Lambda=230 MeV$ (correctly tuned model could give correct pion mass and the right value of  $f_\pi$ in the addition).  Note that the definition of the kernels are chosen in a way it provides $K(0)/\Lambda\simeq 1$ for the dynamical mass $M(\Lambda)\simeq \Lambda$ in absolute value, for instance $K_H(0)=0.88 \Lambda^{-1}$.

\begin{figure} 
\centerline{\epsfig{figure=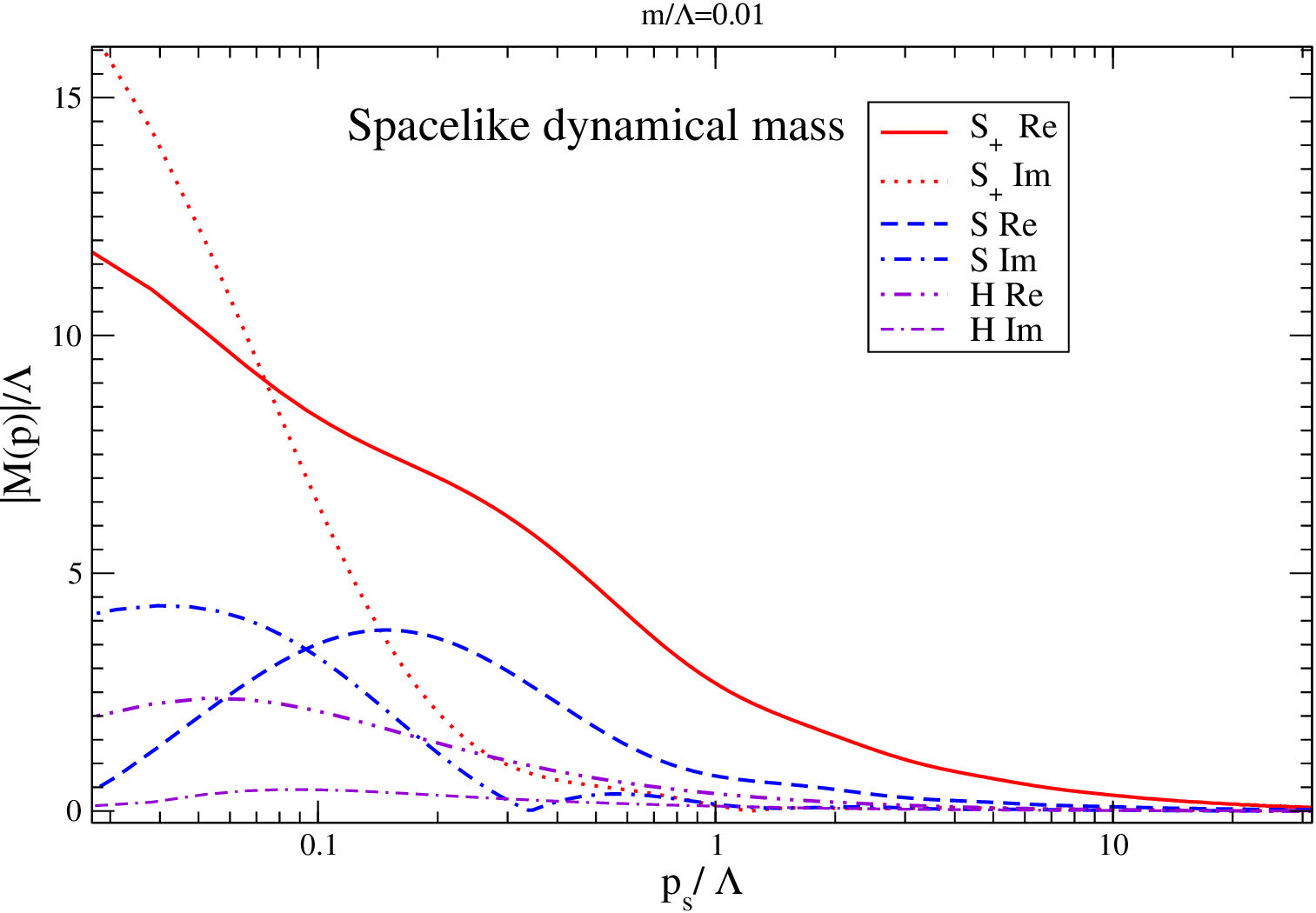,width=7truecm,height=7truecm,angle=0}}
\caption[caption]{\label{first5}Dynamical quark masses as described in the text}
\end{figure} 
\begin{figure} 
\centerline{\epsfig{figure=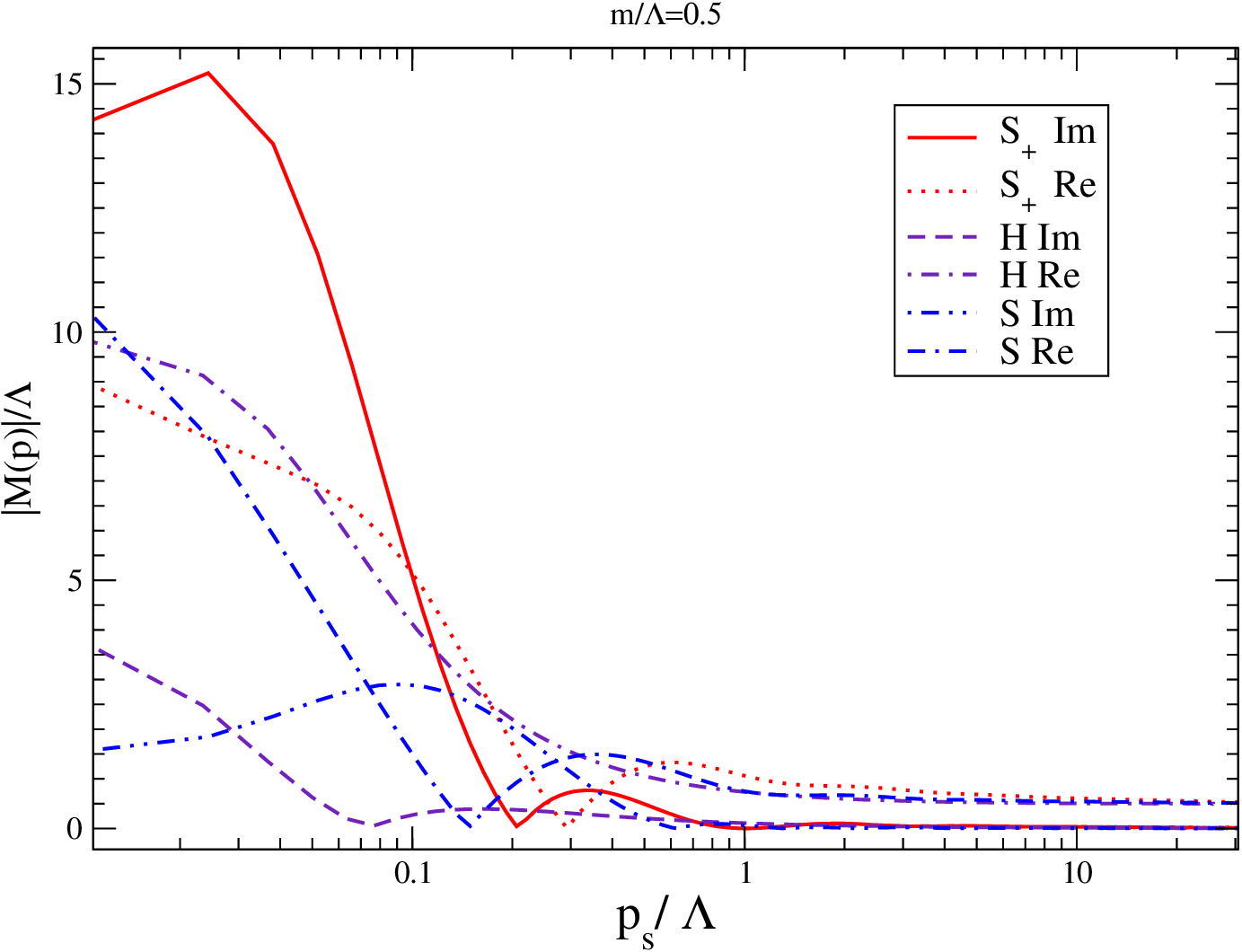,width=7truecm,height=7truecm,angle=0}}
\caption[caption]{\label{first6}Dynamical quark masses as described in the text}
\end{figure}
\begin{figure} 
\centerline{\epsfig{figure=first.eps,width=7truecm,height=7truecm,angle=0}}
\caption[caption]{\label{first}Dynamical quark masses as described in the text}
\end{figure}
\begin{figure} 
\centerline{\epsfig{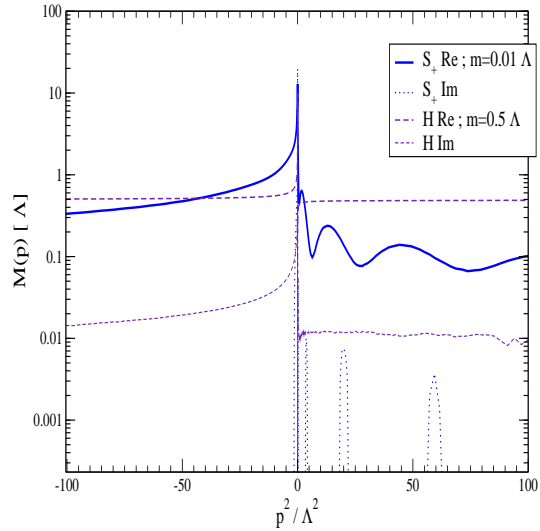}}
\caption[caption]{\label{first4}Dynamical quark masses as described in the text}
\end{figure}

The whole Minkowski space behavior of the dynamical quark mass function is displayed in the Fig \ref{first} and in Fig. \ref{first4}
for log scaled axis of M. The imaginary part is oscillating at the timelike domain of $p^2$ irrespective of the  current quark mass, while it is more smoothed for the kernel with purely meromorphic function (H). We show only two quite different solutions for better visibility, noting that only positive pieces of the absorptive parts are visible for the solution with $K_S$.

In all presented cases it is obvious that  the quark propagator receive nonzero absorptive part everywhere, albeit
it is   decreasing function and  basically turn off at the spacelike scale of several $\Lambda$.  We see the indication that there is a branch point located in the vicinity of zero. Although we are not able identify the appropriate integral weights quantitatively, from the check-list of scalar correlation functions made in the previous Section we expect the infrared part of quark propagator is driven also by the function from $H_+$ and/or $H_-$ spaces. Of course, it must gradually move to the function driven by part from $S^+$ for a high spacelike momenta as it is dictated by QCD perturbative unitarity.        
By studying these and  other cases of $H,S$ functional combinations defining the kernel, we  see the evidence that the resulting quark propagator  is  not entirely  from $S^+$ space, unless the interaction strength (coupling $C$) is very small producing thus infrared quark mass much smaller then $\Lambda$ scale. In this case mass-shell physical pole appears in the propagator.

Recall the solution of DSE in the Euclidean space must be real.
To this point we should stress that it is easy to switch off the imaginary part of the quark propagator at spacelike region
and solve the DSE numerically with ignoring the absorptive part at the spacelike region.
We did it in all studied cases and we actually get the real solution there, which is even stable and convergent for all cases considered,  
however it has never provided  trivial  absorptive spacelike part of $M$ as an output, which is easy to see by the direct evaluation.
We actually did not see any evidence for the solution which  provides chiral symmetry breaking solution and is simultaneously purely real 
at the spacelike. One can  only say  that the absorptive part is rapidly vanishing  at the scale of several $\Lambda$.    
% possible figure first7.agr, but not necessary if nobody wants to see it.
 
\section{Educated Minkowski space  solution  motivated by Euclidean lattice results}

In the previous Section we have presented numerical results for the ladder-rainbow DSE with exponentially suppressed
ultraviolet spacelike modes.The kernel with complex conjugated poles as well as the one based on 
the conventional analytical assumption were used, noting that the all cases lead to a conjecture that quark propagator involves a part consistent with Hilbert transformation providing thus nontrivial imaginary part for the quark propagator at spacelike region. 

According to the recent  findings in Landau gauge, a nontrivial  dressing of quark-gluon vertex is required for a correct description of 
chiral symmetry breaking  via quark DSE \cite{BHATAN2004,ABICAPA2013}. 
In this section we continue a numerical study by using a kernel,  which is more or less motivated by a recent lattice funding.
In accordance with previous argumentation we assume  to quark-gluon vertex  can be complex as well as the quark self energy was.  
 
Let us  remind the   DSE for quark-gluon vertex for this purpose, which reads 
\be \label{vertex}
\Gamma(q,p,k)=\Gamma^{[0]} +ig\int\frac{d^4k}{(2\pi^4)}\Gamma S S V_{qqqq} + ig\int\frac{d^4k}{(2\pi^4)}\Gamma_{ggg} G G V_{qqgg} \, , 
\ee
where $V_{qqqq}$ and $V_{ggqq}$ is a shorthand notation for four-(anti)quarks and 2gluons-2quarks scattering kernels.
Note the later one involves ghost-quark scattering kernel, which according to Euclidean studies should be responsible for
a main enhancement of chiral symmetry breaking effect in the quark DSE.

For any  approximation of the vertex the Minkowski space quark gap equation can be written like
\be \label{maingap}
M(p)= i\int\frac{d^4k}{(2\pi^4)} \frac{M(k)}{k^2-M(k^2)}K(q,p,k)
\ee
where we have defined scalar function $K(q,p,k)$,  which arises after the summation and projection of the product of the quark DSE kernel. Note trivially  that the definition includes the renormalization wave function $A$ as well, however it does not include the boundary condition necessary for a given  analytical continuation, i.e. one should add $\ep$ part in a case of a real pole occurrence.

From Eq. (\ref{maingap}) it is obvious that one does not need to consider all  twelve form factors associated with a various tensorial structure of the quark-gluon vertex when study quark DSE alone. Their role is prominent for calculating of meson properties with various spin. 
The only necessary hint is that all  these form factors lead to the large infrared enhancement of the quark DSE kernel. For an actual quantitative results for a various form factor calculated in the Euclidean space, see the original papers 
\cite{BHATAN2004,ABICAPA2013,HOWIAL2013}. To achieve similar effect, we will make an Ansatz for the kernel $K(q,p,k)$  in the following  way:
\bea \label{kernel}
K(q,p,k)&=& \tilde{C} \tilde{\Gamma}_I(q^2) D(q^2)
\nn \\
\tilde{\Gamma}_I(q^2)&=&\frac{1}{(4\pi)^2}\frac{1}{q^2}Ln{(1-q^2/M^2)}
\nn \\
&=&\frac{1}{2 q^2 (4\pi)^2}\ln{(\frac{(q^2-\Lambda^2)^2+\Lambda^4}{2\Lambda^4})}
\nn \\
&+&\frac{i}{q^2 (4\pi)^2}{\mbox{arctg}}\left(\frac{q^2 }{(q^2-2 \Lambda^2)}\right)  \, ,
\nn \\
D(q)&=&\frac{q^2-i \Lambda^2}{(q^2-\Lambda^2)^2+\Lambda^4}  \, ,
\eea
%codes: 68,66 okj
% and figures quark11.agr and quark12.agr
Where the function $D$ is the  lattice fit -the Gribov gluon propagator, 
however here with slightly more general complex prefactor. The constant  $\Lambda\simeq \Lambda_{QCD}$ is a single dimensionfull parameter of the model.

The choice of the kernel is motivated by an expected appearance of a Lorentz covariant generalization of the linear interquark  potential.  It is chosen in a way that it exactly corresponds  with  the scalar triangle approximation of the  quark gluon vertex function wherein one inner vertex has a vanishing external momentum and wherein two associated connected internal lines have vanishing masses.  In this way the function $\tilde{\Gamma}$  corresponds to the 1 loop convolution of $1/k^4$ function with another "massive" gluon propagator. 
According to Ward identities and our experience with the Minkowski space solutions  studied in the previous section, we expect the quark -gluon vertex 
turns to be complex valued function in the infrared as well. As we are not able to solve Minkowski space quark-gluon DSE at recent  stage, we do this by taking $M^2\rightarrow (\Lambda^2+i\Lambda^2)$ in the kernel.

To solve such quark DSE in Minkowski space is not an easy task and we provide some further technical details in the Appendix B.
The results are shown in Fig. \ref{main1} and in the Fig. \ref{main2} for several values of quark-gluon effective coupling strengths C.
Recall here that at the scale $\Lambda$ the both -real and imaginary- components of quark mass are generated with comparable sizes.  
This should be viewed as a nontrivial consequence of what actually happen when one  assumed complex branch points  
\cite{BHATAN2004,MATA1999} in Minkowski space instead of in the Euclidean one.
It is an another interpretation of $\Lambda_{QCD}$, which is not only the scale where QCD coupling $\alpha_S$ becomes strong but it is the single  parameter which also corresponds with the inverse four-vector distance over which the dressed quark and gluons may propagate before losing its identity by absorption/annihilation in hadronization process.

\begin{figure} 
\centerline{\epsfig{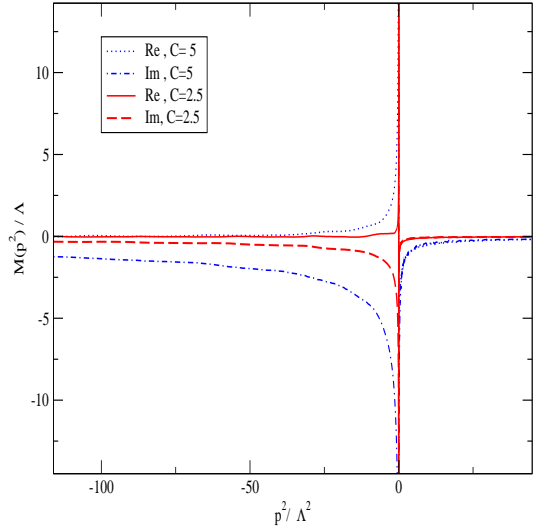}}
\caption[caption]{\label{main1}Dynamical quark mass calculated in a "lattice inspired model" for  various effective coupling $C$.}
\end{figure}

\begin{figure} 
\centerline{\epsfig{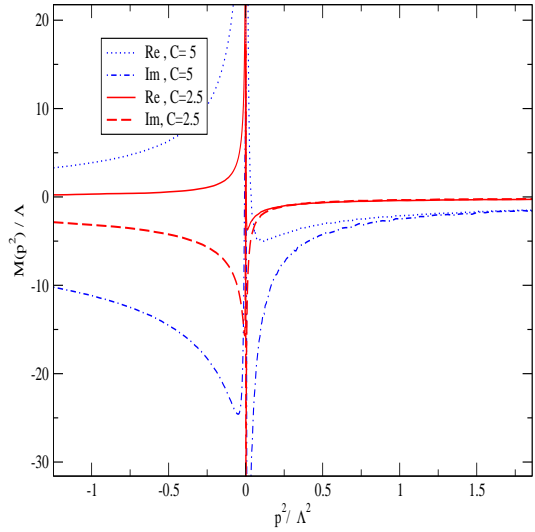}}
\caption[caption]{\label{main2}Dynamical quark mass calculated in a "lattice inspired model" for  various effective coupling $C$ (low $q^2$ view). }
\end{figure}

\section{Conclusion and further prospects}

Motivated by  some recent lattice fit \cite{GRA2010,OLBI2011,DUGU2011,GRA2012,DOS2013} for quark and gluon propagators we have discussed analytical properties of the  scalar correlator. A properties of Stieltjes and Hilbert transformation with and without $\ep$ prescription were used to classify 
each cases separately. Resulting correlator is manifestly Lorentz invariant and it can be written in the dispersion relation with the lowest 
branch point bounded by a positive value even if the propagators have complex conjugated poles. The sufficient condition is that the propagators in correlator are Stieltjes transformable. However, the correlator becomes complex in the spacelike region, if one or both  propagators satisfy Hilbert transformation.  Admitting  QCD correlations function in the form  of Gribov or Cauchy distribution, they should be  complex valued when evaluated in  Minkowski space. In this respect the oldfashionable idea of quark-hadron duality is violated.

In the rest of the paper the  gap equation solutions were presented for the quark propagator employing several approximations of the kernel.
As a warm up a simple kernels with different analytical properties were considered. We end up with the Minkowski study of the lattice inspired model.
It is mere of the fact that for the  interaction strong enough, the dynamical symmetry breaking  is accompanied by generation of complex mass function. 
The real pole is absent as a natural consequence of complex  dynamical mass in agreement with confinement. 

It is needless to say that  avoiding  a numerical  ultraviolet regulators  is impossible if one would like to use the propagators
with a correct perturbative ultraviolet asymptotic. Also, how to include analytical boundary conditions into the more complicated Minkowski space SDE
systems, is recently not obvious to the author. The author is aware that the solution of quark SDE in Minkowski space was actually possible
due to the simplicity of quark DSE, especially since we have analytically well defined DSE kernel and  the quark propagator was the only unknown. 
We expect the same numerical method is plainly not working when two or more unknown functions is under the numerical search (the most urgent could be system of  SDEs for  the gluonic polarization function). We expect that any new and successful method of solution would require an analytical boundary conditions. In other words, it should be a human decision again,  which class of the functional should be chosen for the treatment of DSEs system in Minkowski space. Educated "analytical" continuation of Euclidean data onto the non-analytical cuts of physical momenta is always possible, although the success of the procedure is quite unlike for the function which are oscillating in  Minkowski space-time.

In modern approaches to the bound state problem in QCD, such us DSEs, bound states are built from the propagators of their constituents.
Hadrons therefore share and exhibit the features realized in the quarks propagators, albeit some of these are washout since the constituents propagators
 are always integrated over in various hadronic (Bethe-Salpeter, Fadeev,...) kernels. On of the additional  core of  performance in the first section 
 is to find a new possibility to transform  the original Minkowski momentum space DSEs into a slightly different integral equations,
which will be numerically tractable  in the entire Minkowski space. This program, which very likely requires to go beyond a naive use of the function from $S^+$ space, remains to be done.

Actually, using the symmetry preserving truncation of DSE system the system of Ward-Takahashi identities between the QCD GFS is expected to be valid in both formulation of the theory: Euclidean and Minkowski as well. Let us consider the pion, which is described by homogeneous Bethe-Salpeter equation.    For instance the following well known quark-level Goldberger-Treiman relation \cite{MAROTA1998}:
\be 
f_{\pi}\Gamma_A(k,0)=B(k^2)
\ee
remains valid in the chiral limit in  Minkowski space as well. 
The only difference now, is that the Bethe-Salpeter vertex $\Gamma$ and the pion decay constant
\be 
f_{\pi} P_{\mu}=Z_2Tr_{CD}i\int \frac{d^4k}{(2\pi)^4}\gamma_5\gamma_{\mu}S_f(k+P/2)\Gamma(k,P)S_g(k-P/2)
\ee 
are complex functions in a way that their product reproduces the phase of the complex quark function $B$ correctly.

There are further simple questions I believe can be easily answered in the future.
Up to now, our world is well described by the Standard model and gravity. The Standard Model has field content where only QCD parts exhibit confinement,    the other  fields do not - they are the leptons and electroweak gauge vector bosons. These two sectors  interact together already at classical level since the quarks are charged under the electroweak group. Consequently, at quantum level there are always contributions that unavoidably mix the GFs of confined fields with unconfined ones together.   More concretely, the quark loops, which  contribute to the photon vacuum polarization, must leave the  photon freely propagating. It remains to be shown by an explicit calculation that the quark propagator of the form
\ref{STE2} does not lead to any unexpected badness in the resorts where perturbation theory and usual dispersion theory already successfully works.  We leave these for a future performance.

Within a new precise measurement  of the  pion electromagnetic form factor at the upgraded JLab facility \cite{DUDEK2012}, there is renewed effort
\cite{CCRST2013} in calculation of  pion form factor at low spacelike $Q^2$, e.g.  including  the domain where the perturbative QCD fails. 
It is notable, the recent QCD calculation  has been achieved with a complex  conjugated poles for representation of quark propagator \cite{BPT2003}. 
It would be challenging to perform the similar calculations directly in Minkowski space, where the $\rho$-meson peak  should be automatically generated in quark-photon vertex for timelike photon momenta $q^2$. 

The concept of complex confinement should allow the calculation for excited hadrons as well.
The fail in  achievement the consistent description of highly excited states is  longstanding trouble of QCD practitioners, the habit is to use nonrelativic quantum mechanical description for this purpose \cite{BCPS1997,BBS2001,NB2011}.  In this respect suggested Minkowski space tools can be partially useful when one goes beyond  light-front relativistic quark models description 
usually used for calculation of transition that includes excited nucleons as well \cite{AZBU2012,AZNA2013}.
To this point, in Ref. \cite{SAULI2012} there exists a first reliable Minkowski space calculation of the excited charmonium spectra based on the use of Bethe-Salpeter equation with complex valued charm quark propagator.  The  numerics we have described in presented paper has been actually used  for this purpose, avoiding thus well known problem of Euclidean- Minkowski space continuation for the first time for the Bethe-Salpeter equation.

\section{Appendix A: Assorted integrals}

The spectral representation for the function $F_+$ reads

\be
F_+(q^2)=\int_0^{\infty}\sin{\sqrt{\frac{y}{\Lambda^2}}}\frac{dy}{q^2-y+\ep} \, , 
\ee
which for spacelike $\sqrt{-q^2\Lambda^2}=x$ reads
\be
F_+(q^2)=-\pi e^{-x} 
\ee
and for timelike argument $t=\sqrt{q^2\Lambda^2}$ gives
\be
F_+(q^2)=-\pi \cos{t}-i \pi \sin{t}
\ee

Spectral representation for the kernel $K_{S+}$ reads

\bea
K(q^2)&=&\int_0^{\infty} \rho_{K}(y)\frac{dy}{q^2-y+\ep} \, , 
\nn \\
\rho_{K}(y)&=&Konst \frac{\pi\Lambda^{-2}}{\lambda^2+\Lambda^2}
\left[\cos\left(\frac{\sqrt{y}}{\Lambda}\right)\lambda\Theta(y-\Lambda^2)
+\sin\left(\frac{\sqrt{y}}{\Lambda}\right)\left(-\frac{1}{2}+\frac{\lambda}{\pi}\ln{|\lambda|}\right)\right]
\eea
where $\lambda=1-y/\Lambda^2$ and where we have used the well known algebraic identity \cite{SAU1} valid for the product of two 
functions from $S^+$.

\section{Appendix B: Numerical method}

The lattice theory is without any doubt the most prominent  tool recently used for nonperturbative study of  Quantum Field Theory.
By the construction, it cannot provide complex valued GFS and form factors for purely Euclidean (spacelike) arguments. 
Discretized Minkowski space  analogue of lattice theory is known to be unpractical and according to author knowledge, has never led to a data harvest.  

Also  trying to solve DSEs in Minkowski space is always accomplished  with obstacles related with a number of un-proper principal value integration over the $k_o$ variable.
Let us recall trivial fact, that considering loop integral with perturbative propagators without Feynman $\ep$ prescription
must give trivial result. This is valid for the convolution of any function from the $S$-space discussed in the first section and it  should be  true even for the integrals which are ultraviolet divergent otherwise.  In practice, a numerical subtraction of several  opposite sign infinite pieces is never performed with high accuracy. In case of the DSEs, it usually makes the iteration procedure  never-relaxing when the propagators with usual 
high momentum behavior is considered and certainly, many non-perturbative  QCD calculations represent  ill defined numerical problem in Minkowski space,
Nevertheless this fact  should not lead  to a headlong conclusions  that our real world, i.e. Minkowski space solutions, should be read from the analytical continuations of counterpartners obtained in the Euclidean space (for a traditional discussion see \cite{GJ1987,ROBWIL1994}).

In this Appendix we review the method which is working at least in the cases of our simple models.
For the numerical solutions we  have  chosen the method of iteration.
 As the iterations do not automatically relax for given numerics  the author keep all the (convergent) numerical codes public in the following url \cite{SAULIURL}. In order to avoid numerically uncovenient interpolations we convert gap equation into the form

\be \label{baf}
M(p^2)= \frac{i}{8\pi^3} \int_0^{\infty} dk_0\int_{-\infty}^{k_o^2} d k^2\int_{-1}^{1} dz \sqrt{k_o^2-k^2} K(q^2,p^2,k^2) \frac{M(k^2)}{k^2-M^2(k^2)} \, ;
\ee
with $q^2=k^2+p^2+2\sqrt{k_o^2-k^2}\sqrt{-p^2}z \,$ for a spacelike argument$ p^2<0$, where obvious substitution $\vec{k}\rightarrow k^2$ has been made. 
The  Eq. (\ref{baf}) is 3dim integral equation for the spacelike argument, where variable $z$ is a cosine between $\vec{k}$ and $\vec{p}=(0,0,p_z)$),
For a timelike arguments we are free to choose $p=(p_0,0,0,0)$ for, which allows to write

\bea \label{haf}
M(p^2)&=& \frac{i}{8\pi^3} \sum_{\pm} \int_0^{\infty} dk_0\int_{-\infty}^{k_o^2} dk^2 \sqrt{k_o^2-k^2} K(q^2_{\pm},p^2,k^2) \frac{M(k^2)}{k^2-M^2(k^2)} \, ;
\nn \\
q^2_{\pm}&=&k^2+p^2\pm 2k_o \sqrt{p^2} \, \, {\mbox{for}} \, \, p_0^2=p^2>0.  
\eea
and the DSE is reduced to 2-dim for the timelike external argument in  Eq.(\ref{haf}). We also reduced the number of integration points by a simple substitution $k_o\rightarrow -k_o$ for a negative $k_o$.  For zero $p^2$, the both equations above become identical for the Lorentz invariant mass function $M(p^2)$. The quark mass function $M$ is the function of single scalar  $k^2$ in the kernel.  

We have discretized $k^2$ by using Gaussian method and we have checked our Gaussian integrator in the case 
of complex mass valued Feynman integral which is known exactly. The following integral 
\be \label{lucy}
i\int\frac{d^4l}{(2\pi)^4}\frac{1}{(l^2-a+ib)^3}=\frac{1}{2(4\pi)^2}\frac{a+ib}{a^2+b^2},   
\ee
was actually used for this purpose. Note that the Eq. (\ref{lucy}) corresponds to the scalar triangle with all zero external momenta, wherein all the internal lines correspond to a propagators with the  common single complex valued mass $m=\sqrt(a-ib)$.  Assuming the real numbers $a,b$ satisfy condition $a,b>0$, it allows to perform usual  Wick rotation and analytical integration. Using also the numerical integration  in $Eq.(\ref{haf})$ explicitly we have  reproduced rhs.  of Eq. (\ref{lucy}) within a few promile accuracy.

We have check the method is working for more general kernels than have been presented in the paper here, however the offer for more complicated cases is very limited. For instance, we did not find stable and precise solution to the strong coupling models where both propagators are unknown in the approximated expressions for selfenergy (we considered a simple version of cubic and Yukawa models).    
To get the solution a real poles must be avoided, i.e. if an effective  coupling considered is too weak the confinement is lost and we usually get a trivial or oscillating unstable solution. The singularity of propagator at the real axis is the  avoided for chirality breaking nontrivial solution only.
Furthermore, a nodal solutions were obtained in some cases. At least three solutions exist at chiral symmetry breaking phase with zero current quark mass: one which is trivial and the other two differ by the sign. We presented the results as they have been obtained numerically, i.e without changing the sign in any case.

The concept of analytical confinement is actually not new, however there is only quite indirect evidence of it,  when the Euclidean metric is used  as definite one (see \cite{BHATAN2004,MATA1999,ATK1979,MAHO1991,MARISQED2+1,GILL2008} for related topics). We argue it is related with the complexity of QCD GFs, which however does not contradict any of basic principles of quantum field theory, e.g. unitarity,  reality of energy spectra, etc. since  this complexification is own property of  GFs of confined objects - those that do not exhibit themselves  as a real poles and henceforth 
do not appear in the asymptotic states of S-matrix.


\begin{thebibliography}{99}
\bibitem{SAULIURL}
The codes are available at the following url: gemma.ujf.cas.cz/~sauli 
from the date the paper is accepted for publication.
\end{thebibliography}
\end{document}